\documentclass[twocolumn,showpacs,preprintnumbers]{revtex4}
\usepackage{amssymb}
\usepackage[dvips]{graphicx}

\begin{document}

\title{Small adiabatic polaron with a long-range electron-phonon interaction}

\author{A.S.Alexandrov$^1$ and B.Ya.Yavidov$^{1,2}$}

\address{$^1$Department of Physics, Loughborough University, Leicestershire, LE11 3TU UK\\
$^{2}$Nukus State Pedagogical Institute named after A'jiniyaz, 742005 Nukus,
        Karakalpakstan, Uzbekistan}

\begin{abstract}
Two-site single electron system interacting with many vibrating
ions of a lattice via a long-range (Fr\"{o}hlich) electron-phonon
interaction is studied in the adiabatic regime. The renormalised
hopping integral of small adiabatic Fr\"{o}hlich polarons is
calculated and compared with the hopping integral of small
adiabatic Holstein polarons.

\end{abstract}

\maketitle

 If phonon frequencies are very low, the
local lattice deformation due to the electron-phonon interaction
can trap the electron. This {\it self-trapping }phenomenon was
predicted by Landau \cite{lan2}. It has been studied in greater
detail by Pekar \cite{pek}, Fr\"{o}hlich \cite{fro2}, Feynman
\cite{fey0}, Devreese \cite{dev} and other authors in the
effective mass approximation, which leads to a so-called {\it
large polaron.}

When the electron-phonon coupling  is relatively strong, $\lambda=E_{p}/D
>1$, all electrons in the Bloch band are "dressed" by phonons. In this regime the electron kinetic energy, which is less than the half-bandwidth ($D$), is small compared with the potential energy due to a local lattice deformation, $E_{p}$, caused by electrons themselves.  Here the finite bandwidth is essential, and the effective mass approximation cannot be applied. The electron is called a {\it small polaron }in this regime.  The main features of small polarons were understood by Tjablikov \cite{tja}, Yamashita and Kurosava \cite{yam}, Sewell \cite{sew}, Holstein \cite{hol} and his school \cite{frihol,emi}, Lang and Firsov \cite{fir}, Eagles \cite{eag}, and others and described in several review papers and textbooks \cite{app,fir2,dev,bry,mah,alemot}.  In particular, in his pioneering study of the small polaron dynamics Holstein \cite{hol} introduced a two-site molecular model of a single electron coupled with local (molecular vibrations). He derived a renormalised (polaron) hopping integral for two extreme cases, the nonadibatic, $\omega \geq t$ and adiabatic, $\omega \ll t$, where $t$ is the bare (unrenormalised) hopping integral, and $\omega $ is the characteristic phonon energy. An exponential reduction of the bandwidth at large values of $\lambda $ is a hallmark of  small polarons. The self-trapping is never ``complete'' even in the strong-coupling regime, that is any polaron can tunnel through the lattice. Only in the extreme {\it adiabatic} limit, when the phonon frequencies tend to zero, the self-trapping is complete, and the polaron motion is no longer translationally continuous.

\vspace{0.3 cm}

In the last years quite a few numerical and advanced analytical studies  confirmed these and subsequent \cite{fir} analytical results (see, for example, \cite{gog,alePRB,kab,kab2,bis0,mar,tak,feh,tak2,rom,lam,zey,wag,tru,alekor,aub2,ale00}).
At the same time polarons were experimentally recognised as quasiparticles in novel materials, in particular, in superconducting cuprates and colossal magnetoresistance manganites \cite{exp}.

\vspace{0.3 cm}

Small polarons are  very heavy due a local character of the electron-phonon interaction  in the Holstein model, so that their huge effective mass created some prejudices with respect to any relevance of polarons to real oxides. However, it was pointed out that  small polarons could be quite mobile  if one takes into account a more realistic long-range interaction with phonons in ionic solids \cite{asa}. Indeed the Monte-Carlo \cite{alekor} and other calculations \cite{bt,flw} proved that small polarons with a long-range Fr\"{o}hlich interaction are a few orders of magnitude lighter than the small Holstein  polarons (SHP) with the same binding energy. All numerical and analytical results for these, so-called small  Fr\"{o}hlich  polarons (SFP) \cite{alekor}, were obtained in the nonadiabatic or near-nonadibatic regimes. Here we extend these
studies to the adiabatic region.

We consider an extended Holstein model of an electron hopping
between two sites, but interacting with all surrounding ions of
the lattice via a long-range electron-phonon coupling, like in a simple case of a one-dimensional ionic chain, vibrating in the direction, perpendicular to the chain, as shown in Fig.1.
\begin{figure}[tbp]
\begin{center}
\includegraphics[angle=-0,width=0.3\textwidth]{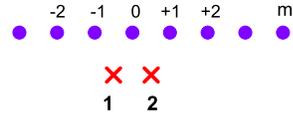} \vskip -0.5mm
\end{center}
\caption{Electron interacting with infinite chain of ions via
long-range  interaction performing a hop between two sites {\textbf{1}}
and {\textbf{2}}.}
\end{figure}
\vspace{0.5cm}

For simplicity, sites {\bf 1} and {\bf 2} are  not
vibrating, but the interaction with their vibrations can be easily
included in our model.  The model mimics  $high-T_{c}$ cuprates,
where in-plane ($CuO_{2}$) carriers are strongly coupled with the
$c$-axis polarized vibrations of  $apical$ oxygen ions \cite{t}.
We derive an analytical expression for the renormalised hopping
integral  and compare it with SHP in the adiabatic limit.

The Hamiltonian of the model  is
\begin{equation}
H=H_{ph}+H_{e}+H_{e-ph}
\end{equation}
where
\begin{equation}
H_{ph}=\sum_{\bf m}\left(-\frac{\partial^2}{2M\partial u^{2}_m}+\frac{M\omega^2 u^{2}_m}{2}\right)
\end{equation}
is the Hamiltonian of vibrating ions,
\begin{equation}
H_{e-ph}=\sum_{\bf i=1,2}\sum_mc^{\dagger}_{\bf i}c_{\bf i}f_m({\bf i})u_m
\end{equation}
describes interaction between the electron and the ions, and $H_{e}=-t (c^{\dagger}_{\bf 1}c_{\bf 2}
+H.c.)$  is the electron hopping energy. Here $u_m$ is the
displacement and $f_{m}({\bf i})$ is an interacting
force between electron on site {\bf i} and the m-th ion. $M$ is the mass of vibrating ions
and $\omega$ is their frequency.

The wave function is a superposition of two normalized Wannier
functions $W({\bf r})$, localized on the left and the right sites,
\begin{equation}
\Psi=a_{\bf 1}(u_m)W({\bf r}-{\bf 1})+a_{\bf 2}(u_m)W({\bf r}-{\bf 2}),
\end{equation}
The Schr\"{o}dinger equation is reduced to two coupled
 second order differential equations with respect
to the $infinite$ number of  vibrational coordinates $u_m$
\begin{equation}
\left[E-\sum_m(H_{ph}-f_m({\bf 1})u_m)\right]a_{\bf 1}(u_m)=ta_{\bf 2}(u_m)
\end{equation}
\begin{equation}
\left[E-\sum_m(H_{ph}-f_m({\bf 2})u_m)\right]a_{\bf 2}(u_m)=ta_{\bf 1}(u_m)
\end{equation}
Let us first discuss the nonadiabtic regime. Following Holstein
\cite{hol} we apply a perturbation approach with respect to the
hopping integral. In the zero order ($t=0$) the electron is
localized either on the left or on the right site, so that
\begin{equation}
a^{l}_{\bf 1}(u_m)=\exp \left[-\frac{M\omega}{2}\sum_m\left(u_m+\frac{f_m({\bf 1})}{M\omega^2}\right)^2\right], a^{l}_{\bf 2}(u_m)=0
\end{equation}

or

\begin{equation}
a^{r}_{\bf 1}(u_m)=0, a^{r}_{\bf 2}(u_m)=\exp\left[-\frac{M\omega}{2}\sum_m\left(u_m+\frac{f_m({\bf 2})}{M\omega^2}\right)^2\right]
\end{equation}

In the first order  $a_{\bf 1}(u_m)$ and
$a_{\bf 2}(u_m)$ are linear combinations of
$a^{l}_{\bf 1}(u_m)$ and $a^{r}_{\bf 2}(u_m)$ as

\begin{equation}
\left(%
\begin{array}{c}
  a_{\bf 1}(u_m)\\
  a_{\bf 2}(u_m)\\
\end{array}%
\right)=
\alpha
\left(%
\begin{array}{c}
  a^{l}_{\bf 1}(u_m) \\
  0 \\
\end{array}%
\right)
+\beta
\left(%
\begin{array}{c}
  0 \\
  a^{r}_{\bf 2}(u_m) \\
\end{array}%
\right).
\end{equation}

Substituting Eq.(9) into Eq.(5) and Eq.(6) one gets a system of linear equations with respect
to ${\alpha}$ and ${\beta}$. The standard \cite{hol} secular equation for the energy is
\begin{equation}
det\left(%
\begin{array}{cc}
  E-N\omega/2+E_{p} & \widetilde{t} \\
  \widetilde{t} & E-N\omega/2+E_{p} \\
\end{array}%
\right)=0
\end{equation}
where $E_{p}=\sum_mf_m^2({\bf 1})/2M\omega^2$ is the polaronic shift, $N$ is the number of ions in the chain and
\begin{equation}
\widetilde{t}=t\left[\int
a^{l}_{\bf 1}(u_m)a^{r}_{\bf 2}(u_m)du_m/\int
|a^{l}_{\bf 1}(u_m)|^{2}du_m\right]
\end{equation}
is a renormalised hopping integral. Then the lowest energy levels of the system are found as $E_{\pm}=N\omega/2-E_{p}\pm\widetilde{t}$. The evaluation of Eq.(11) with explicit
$a^{l}_{\bf 1}(u_m)$ and $a^{r}_{\bf 2}(u_m)$
results in $\widetilde{t}=t\exp \left(-g^{2}\right)$, where
\begin{equation}
g^{2}=\frac {1}{2M\omega^3} \sum_m\left(f^{2}_m({\bf 1})-f_m({\bf 1})f_m({\bf 2})
\right). 
\end{equation}

The same expression for polaronic shift was obtained in Ref. \cite{alekor} by using the canonical Lang-Firsov transformation. We would like to stress that the model yields less renormalization of the effective mass than the Holstein model. For simplicity let us take into account only nearest-neighbors interactions, as Fig.2{\it a}. In this case our model yields $E_p=f^{2}_{\bf 0}({\bf 1})/M\omega^2$ and the mass renormalization $m^{\ast}/m=\exp(E_{p}/2\omega)$, while the Holstein model with the local interaction, Fig.2{\it b}, yields $m^{\ast}/m=\exp(E_{p}/\omega)$.

\begin{figure}[tbp]
\begin{center}
\includegraphics[angle=-0,width=0.4\textwidth]{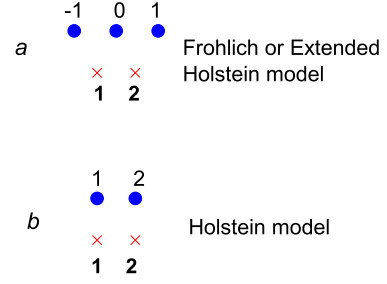} \vskip -0.5mm
\end{center}
\caption{Schematic representation of the Fr\"{o}hlich (or Extended Holstein) model and the Holstein model. The electron on the site ${\bf 1}$ interacts ({\it a}) with the sites $m=-1$ and $m=0$ and ({\it b}) with only site $m=1$ of the ion chain,  in the Fr\"{o}hlich  and the Holstein model, respecively.}
\end{figure}
\vspace{0.5cm}
The factor $1/2$ in the exponent provides much lighter Fr\"{o}hlich small polarons compared with the Holstein model, Fig. 3. If one considers the Coulomb-like interaction with the whole upper chain, one gets the factor 0.28 \cite{trg} instead of 0.5 in the exponent, that means even less renormalised effective mass.

\begin{figure}[tbp]
\begin{center}
\includegraphics[angle=-0,width=0.47\textwidth]{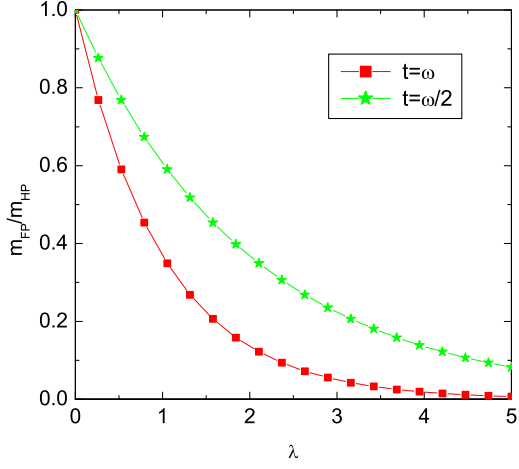} \vskip -0.5mm
\end{center}
\caption{Ratio of masses of nonadiabatic Fr\"{o}hlich and Holstein small polarons as a function of the
electron-phonon coupling constant $\lambda$ at different values of $\omega/t$.}
\end{figure}
In the opposite adiabatic regime  we use the Born-Oppenheimer approximation 
representing the wave function as a product of wave functions describing the "vibrating" ions, $\chi(u_m)$, and the electron with a "frozen" ion displacements, $\left(%
\begin{array}{cc}
  \psi(u_m) & \varphi(u_m) \\
\end{array}%
\right)^{T}$ ($T$ means transpose matrix). Terms with the first and second derivatives of the "electronic" functions $\psi(u_m)$ and $\varphi(u_m)$ are small compared with the corresponding terms with derivatives of $\chi(u_m)$. The wave function of the "frozen" state obeys the following equations

\begin{equation}
\left[E(u_m)-\sum_mf_m({\bf 1})u_m\right]\psi(u_m)-t\varphi(u_m)=0
\end{equation}
\begin{equation}
-t\psi(u_m)+\left[E(u_m)-\sum_mf_m({\bf 2})u_m\right]\varphi(u_m)=0.
\end{equation}
The lowest energy is
\begin{equation}
E(u_m)=\frac{1}{2}\sum_mf^{+}_m u_m-[\frac{1}{4}(\sum_mf^{-}_mu_m)^2+t^2]^{1/2},
\end{equation}
that plays a role of potential energy in the equation for $\chi(u_m)$
\begin{equation}
\left[E-\sum_m H_{ph}-E(u_m)\right]\chi(u_m)=0.
\end{equation}

Here $f^{+}_m=\left[f_m({\bf 1})+f_m({\bf 2})\right]$ and $f^{-}_m=\left[f_m({\bf 1})-f_m({\bf 2})\right]$. Because of the infinite number of variables, Eq.(16) can not be reduced to a double-well potential problem \cite{hol}. However, in the nearest-neighbor approximation with three relevant ions in the upper row, Fig.2{\it a}, it can. Indeed, using the transformation $\xi_m=u_m+u_{-m}$ and $\eta_m=u_m-u_{-m}$ and $f^{+}_m=f^{+}_{-m}$, $f^{-}_m=-f^{-}_{-m}$,  one can rewrite the latter equation as

\begin{equation}
\left(E-2\omega+\frac{3}{4}E_p+\frac{\partial^2}{2\mu\partial\eta^{2}_1}-U(\eta_1)\right)\chi(\eta_1)=0.
\end{equation}
Here
\begin{equation}
U(\eta_1)=\frac{\mu\omega^2 \eta^{2}_1}{2}-\left[\frac{1}{4}\left(f^{-}_1\eta_1\right)^2 +t^2\right]^{1/2}
\end{equation}
is the familiar double-well potential, and $\mu=M/2$.
Then the determination of the energy splitting is similar to the case considered in \cite{hol}. The splitting appears due to penetrable character of the double-well potential:
\begin{equation}
\Delta E=\frac{\widetilde{\omega}}{\pi}{\it exp}\left[-2\int\limits_{0}^{\eta_{tp}}|p({\eta_1})|d{\eta_1}\right],
\end{equation}
where $\eta_{tp}=1-1/\sqrt{\mu\widetilde{\omega}}$  is the classical turning point corresponding to the energy of the ground state $E_{0}=\widetilde{\omega}/2+U_{min}(\eta_1)$ and $p(\eta_1)=\sqrt{2\mu(E_{0}-U(\eta_1))}$ is the classical momentum. The result is $\Delta E=\Delta \exp(-g^{2}_F)$, where
\begin{equation}
\Delta=\frac{\widetilde{\omega}}{\pi}\sqrt{\frac{E_{p}}{2\omega}\kappa^{3/2}}\left(1-\sqrt{1-\left(\frac{E_{p}}{2\omega}\kappa^{3/2}\right)^{-1}}\right),
\end{equation}
and
\begin{equation}
g^{2}_F=\frac{E_{p}}{2\omega}\kappa^{1/2}\sqrt{1-\left(\frac{E_{p}}{2\omega}\kappa^{1/2}\right)^{-1}}.
\end{equation}
Here $\widetilde{\omega}=\omega\sqrt{\kappa}$ is the renormalised phonon frequency, $\kappa=(1-1/\lambda^2)$ and $\lambda=E_p/(2t)$. In the Holstein model one should replace $E_p$ in Eq.(20) and Eq.(21) by $2E_p$. The comparison of two models shows that the polaron in the Fr\"{o}hlich model remains much lighter, than the Holstein polaron also in adiabatic regime, Fig.4.
\begin{figure}[tbp]
\begin{center}
\includegraphics[angle=-0,width=0.47\textwidth]{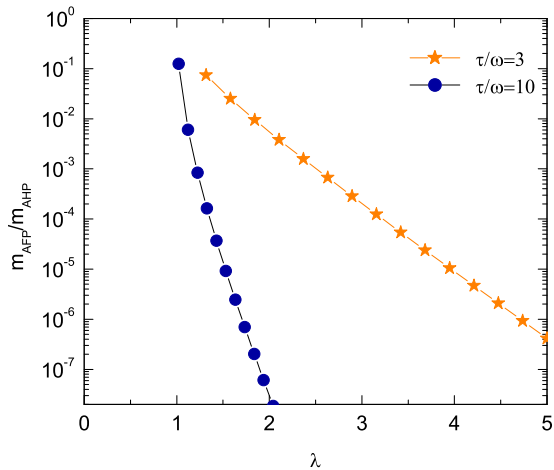} \vskip -0.5mm
\end{center}
\caption{Ratio of the SFP  mass to the SHP mass as a function of ${\lambda}$ in the adiabatic regime.}
\end{figure}

\vspace{ 0.5 cm}

In conclusion, we have solved an extended Holstein model with a
long-range Fr\"{o}hlich interaction in the adiabatic limit. We
have found the  hopping integral of SFP, and compared it with SHP.
The small adiabatic Fr\"{o}hlich polaron is found many orders of magnitude
lighter than the small Holstein polaron both in the nonadiabatic
and adiabatic, Fig.4, regimes. One of us (B.Ya.Ya) is grateful to NATO and
the Royal Society for their financial support (grant PHJ-T3). We also acknowledge the support of Leverhulme Trust, UK.


\begin{thebibliography}{99}
\bibitem{lan2}  {\small L.D.Landau, {\it \ J. Phys.} {\it (USSR)} {\bf 3},
664 (1933).}

\bibitem{pek}  {\small S.I.Pekar, {\it Zh. Eksp. Teor. Fiz.}{\bf \ 16},
335 (1946).}

\bibitem{fro2}  {\small H.Fr\"{o}hlich, {\it \ Adv. Phys. }{\bf 3}, 325 (1954).}

\bibitem{fey0}  {\small R.P.Feynman, {\it Phys. Rev.} {\bf 97}, 660 (1955).}

\bibitem{dev}  {\small J.T.Devreese, in {\it Encyclopedia of Applied
Physics}, edited by G.L.Trigg (VCH Publishers, New York, 1996), Vol. {\bf 14}, p. 383  and references
therein.}

\bibitem{tja}  {\small S.V.Tjablikov, {\it Zh.Eksp.Teor.Fiz.} {\bf 23},
381 (1952).}

\bibitem{yam}  {\small J.Yamashita and T. Kurosawa, {\it \ J. Phys.
Chem. Solids} {\bf 5}, 34 (1958).}

\bibitem{sew}  {\small G.L.Sewell, {\it Phil. Mag}. {\bf 3}, 1361 (1958).}

\bibitem{hol}  {\small T.Holstein, {\it Ann. Phys.} {\bf 8}, 325 (1959); {\it %
ibid} {\bf 8}, 343 (1959).}

\bibitem{frihol}  {\small L.Friedman and T.Holstein, {\it Ann. Phys}. {\bf 21},
494 (1963).}

\bibitem{emi}  {\small D.Emin and T.Holstein, {\it Ann. Phys}. {\bf 53}, 439 (1969).}

\bibitem{fir}  {\small I.G.Lang and Yu.A.Firsov, {\it Zh. Eksp. Teor.
Fiz.} {\bf 43}, 1843 (1962) [{\it Sov. Phys. JETP} {\bf 16}, 1301 (1963)].}

\bibitem{eag}  {\small D.M.Eagles, {\it Phys. Rev.} {\bf 130}, 1381 (1963); {\it %
ibid} {\bf 181}, 1278 (1969); {\it ibid} {\bf 186}, 456 (1969).}

\bibitem{app}  {\small J.Appel, in {\em Solid State Physics},
(eds. F.Seitz, D.Turnbull and H.Ehrenreich (Academic Press, New York, 1968), Vol. {\bf 21}).}

\bibitem{fir2}  {\small {\it Polarons}, edited by Yu.A.Firsov (Nauka, Moscow, 1975).}

\bibitem{bry}  {\small H.B\"{o}ttger and V.V.Bryksin, {\em Hopping
Conduction in Solids} (Academie-Verlag, Berlin, 1985).}

\bibitem{mah}  {\small G.D.Mahan, {\em Many Particle Physics} (Plenum Press, New York, 1990).}

\bibitem{alemot}  {\small A.S. Alexandrov and N.F. Mott, {\it Rep. Prog. Phys.} {\bf\ 57}, 1197
(1994); {\em Polarons and Bipolarons} (World Scientific, Singapore, 1995).}

\bibitem{gog}  {\small A.A.Gogolin, {\it Phys.Status Solidi} B{\bf 109},
95 (1982).}

\bibitem{alePRB}  {\small A.S.Alexandrov, {\it Phys. Rev}. B{\bf 46}, 2838 (1992).}

\bibitem{kab}  {\small V.V.Kabanov and O.Yu.Mashtakov, {\it Phys. Rev.} B{\bf 47}, 6060 (1993).}


\bibitem{kab2}  {\small A.S.Alexandrov, V.V.Kabanov, and D.K.Ray, {\it %
Phys. Rev.} B{\bf 49}, 9915 (1994).}

\bibitem{bis0}  {\small A.R.Bishop and M.I.Salkola, in {\em Polarons and
Bipolarons in High-$T_{c}$ Superconductors and Related Materials} 
edited by E.K.H.Salje, A.S.Alexandrov, and W.Y.Liang (Cambridge University Press, Cambridge, England, 1995), p. 353 }

\bibitem{mar}  {\small F.Marsiglio, Physica C {\bf 244}, 21 (1995).}

\bibitem{tak}  {\small Y.Takada and T.Higuchi, {\it \ Phys. Rev.} B{\bf 52%
}, 12720 (1995).}

\bibitem{feh}  {\small H.Fehske, J.Loos, and G.Wellein, {\it Z. Phys.} B%
{\bf 104}, 619 (1997).}

\bibitem{tak2}  {\small T.Hotta and Y.Takada, {\it Phys. Rev.} B {\bf 56},
13 916 (1997).}

\bibitem{rom}  {\small A.H.Romero, D.W.Brown and K.Lindenberg, {\it J.
Chem. Phys.} {\bf 109}, 6504 (1998). }

\bibitem{lam}  {\small A.La Magna and R.Pucci, {\it Phys. Rev.} B{\bf 53},
8449 (1996).}

\bibitem{zey}  {\small P.Benedetti and R.Zeyher, {\it Phys. Rev.} B{\bf %
58}, 14320 (1998).}

\bibitem{wag}  {\small T.Frank and M.Wagner, {\it Phys. Rev}. B {\bf 60},
3252 (1999).}

\bibitem{tru}  {\small J.Bonca, S.A.Trugman, and I.Batistic, {\it Phys.
Rev.} B{\bf 60}, 1633 (1999).}

\bibitem{alekor}  {\small A.S.Alexandrov and P.E.Kornilovich, {\it \
Phys. Rev. Lett.} {\bf 82}, ll807 (1999).}

\bibitem{aub2}  {\small L.Proville and S.Aubry, {\it Eur. Phys. J.} B
{\bf 11}, 41 (1999).}

\bibitem{ale00}  {\small A.S.Alexandrov, {\it Phys. Rev.} B{\bf 61},
12315 (2000).}

\bibitem{exp}  {\small see for example: D.Mihailovic, C.M.Foster, K.Voss and A.J.Heeger, 
Phys. Rev. B{\bf 42}, 7989 (1990)}; {\small P.Calvani, M.Capizzi, S.Lupi,
P.Maselli, A.Paolone, P.Roy, S-W.Cheong, W.Sadowski and E.Walker
 {\it Solid State Commun.} {\bf 91}, 113 (1994);} {\small G.Zhao, M.B.Hunt
, H.Keller, and K.A.M\"{u}ller, {\it Nature} (London) {\bf
385}, 236 (1997)}; {\small A.Lanzara, P.V.Bogdanov, X.J.Zhou, S.A.Kellar, 
D.L.Feng, E.D.Lu, T.Yoshida, H.Eisaki, A.Fujimori, K.Kishio,
J.I.Shimoyama, T.Noda, S.Uchida, Z.Hussain, Z.X.Shen, {\it
Nature} (London) {\bf 412}, 510 (2001)}; {\small T.Egami, {\it J. Low Temp.
Phys.}{\bf \ 105 }, 791 (1996)}; {\small D.R.Temprano, J.Mesot, S.Janssen
, K.Conder, A.Furrer, H.Mutka, and K.A.M\"{u}ller, {\it Phys.
Rev. Lett.} {\bf 84}, 1990 (2000)}; {\small Z.X.Shen, A.Lanzara
, S.Ishihara, N.Nagaosa, {\it Phil. Mag.} B{\bf 82}, 1349 (2002).}

\bibitem{asa}  {\small A.S.Alexandrov, {\it Phys. Rev.}
B{\bf 53}, 2863 (1996).}

\bibitem{bt}  {\small J.Bonca and S.A.Trugman, {\it Phys. Rev.} B {\bf\ 64}, 094507 (2001).}

\bibitem{flw}  {\small H.Feshke, J.Loos, and G.Wellein, {\it Phys. Rev.} B {\bf\ 61}, 8016 (2000).}

\bibitem{t}    {\small T.Timusk, C.C.Homes, and W.Reichardt, in Anharmonic Properties of High-TC Cuprates
, edited by D.Mihailovic $et$ $al.$ (Wolrd Scientific, Singapore,
1995), p.171}
\bibitem{trg}  {\small S.A.Trugman, J.Bon\v{c}a, and Li-Chung Ku, {\it Int. J. Modern Phys.} B {\bf\ 15}, 2707 (2001).}

\end{thebibliography}
\end{document}